# Machine Learning-Assisted Analysis and Inverse Design of Prism-Based Surface Plasmon Resonance Sensors

R. Runthala, S. Murai, and P. Arora

***Abstract*—In this work, we have demonstrated the use of data-driven machine learning (ML) models for an efficient design of the Kretschmann configuration-based surface plasmon resonance (SPR) sensor. A diverse physics-based dataset was generated using a MATLAB-based transfer matrix method, spanning a wide range of material optical properties, layer thicknesses, and structural configurations relevant to prism-based SPR sensors. Optical characteristics and thickness are the input features, while SPR performance parameters (figure of merit (FOM) and minimum reflectance ($R_{min}$)) are the output features. A broad comparative study is conducted across different ML models: CatBoost, XGBoost, LightGBM, and a Neural Network (MLP). These models are evaluated using key performance metrics, including R-squared ($R^2$), mean absolute error, and root mean squared error. To provide a comprehensive evaluation framework, this work integrates comparative ML benchmarking, SHAP-based explainability, robustness assessment, and optimization-driven inverse design within a unified workflow. SHAP-weighted perturbation experiments are further conducted to evaluate the model's robustness to realistic input perturbations. Four optimization algorithms are employed for inverse sensor design, and the resulting configurations are validated against parameters and performance recovery analysis. The optimizers are further evaluated on a complementary forward design task, in which independent runs converged on a common optimal sensor configuration, with predicted performance confirmed to be within approximately 0.6-0.9% FOM error and under 0.7% $R_{min}$ error relative to TMM simulation. The proposed ML framework achieved excellent predictive performance ($R^2 > 0.999$ for FOM and $R^2$ up to 0.996 for $R_{min}$ across the best-performing models) while reducing computational time from seconds to milliseconds per sensor configuration, corresponding to an acceleration of approximately $10^3$- $10^4$-fold compared with running the MATLAB TMM simulation directly. The results demonstrate that integrating surrogate ML models with optimization algorithms provides an effective, computationally efficient framework for rapid SPR sensor analysis, inverse design, and performance optimization.**

## I. INTRODUCTION

Surface Plasmon Resonance (SPR)-based optical sensors have attracted significant attention due to their label-free, noninvasive, and highly sensitive sensing capabilities. These sensors have found wide-ranging applications in gas sensing, biosensing, medical diagnostics, virus and bacteria detection, and cancer analysis [1]. The exploration of new materials and innovative multilayer device configurations has largely driven the continuous improvement in SPR sensor performance. Researchers are actively investigating materials across every layer of the SPR stack, including prism materials, plasmonic metals, dielectric layers, and, more recently, two-dimensional (2D) nanomaterials.

Advances in computational electromagnetic modeling have considerably accelerated the analysis and design of SPR sensors, which would otherwise require extensive experimental trials. Multilayer SPR structures are commonly analyzed using methods such as the Transfer Matrix Method (TMM), Rigorous Coupled-Wave Analysis (RCWA), and the Finite Element Method (FEM), which are typically implemented in simulation platforms such as MATLAB and COMSOL Multiphysics. However, designing high-performance multilayer SPR sensors remains computationally demanding because each candidate configuration requires repeated electromagnetic simulations across a large parameter space spanning different materials, thicknesses, and structural arrangements.

To quantify the design space complexity of multilayer SPR sensors, the total number of feasible sensor configurations can be expressed as

$$N_{confi} = G\ P \prod_{i=1}^{n} C_i T_i \tag{1}$$

where $G$ denotes the number of glass prism materials, $n$ is the number of functional layers, $C_i$ and $T_i$ represent the numbers of candidate materials and discrete thickness values for the $i^{th}$ functional layer, respectively, and P denotes the number of allowable layer permutations (For a fixed layer order, P =1).

P. A. would like to acknowledge BITS Pilani (CDRF, SPARKLE, AKS-DTRF) for the financial support. (Corresponding author: Dr. P. Arora).

R. Runthala: Conceptualization, methodology, investigation, analysis, and writing original draft. S. Murai: Investigation and writing review and editing.
P. Arora: Supervision, methodology, investigation, and writing review and editing.

R. Runthala and P. Arora are with the Department of Electrical and Electronics Engineering, Birla Institute of Technology & Science Pilani, Rajasthan-333031, India.
(e-mail: p20230044@pilani.bits-pilani.ac.in, pankaj.arora@pilani.bits-pilani.ac.in).
S. Murai is Department of Physics and Electronics, Graduate School of Engineering, Osaka Metropolitan University, B4-W206, 1-1 Gakuen-cho, Naka-ku, Sakai, Osaka 5998531, Japan.
(e-mail: murai@omu.ac.jp).

Equation (1) indicates that the design space scales multiplicatively with the number of material choices, discrete thickness values, and allowable layer permutations. Consequently, even a modest increase in the number of design variables results in a substantial increase in the number of feasible sensor configurations. Figure 1 shows a representative illustration of the combinatorial growth of the prism-based SPR design space. To further illustrate this combinatorial growth, representative multilayer SPR architectures, along with the corresponding numbers of feasible sensor configurations, are summarized in Table I.

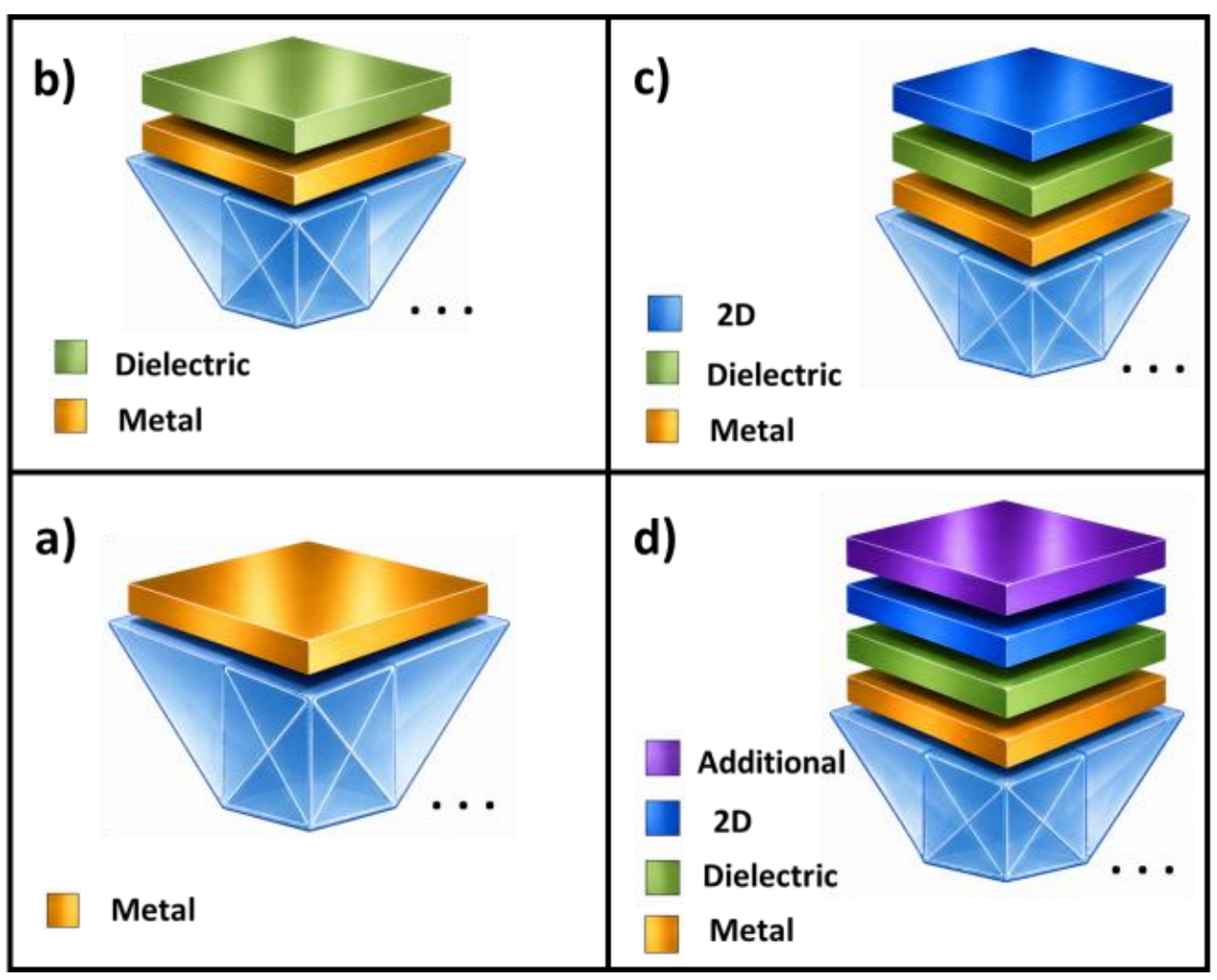

**Fig. 1**. Representative prism-based SPR sensor architectures illustrating the progressive expansion of the design space.

Even with a limited number of candidate materials and discrete thickness values, the search space expands rapidly as additional functional layers are introduced. Furthermore, allowing different layer permutations further multiplies the search space, making exhaustive TMM-based optimization increasingly impractical.

TABLE I. Illustrative combinatorial growth of the SPR design space

| Case | SPR Architecture | Functional Layers | $N_{confi}$ | Increase in Search space |
|---|---|---|---|---|
| (a) | Prism + Metal | 1 | 186 | - |
| (b) | Prism + Metal + Dielectric | 2 | 5580 | ×30 |
| (c) | Prism + Metal + Dielectric + 2D Material | 3 | 167,400 | ×30 |
| (d) | Prism + Metal + Dielectric + 2D Material + Additional Layer | 4 | 5,022,000 | ×30 |

As the search space expands from a few hundred to a million candidate configurations, repeated TMM simulations become computationally expensive and significantly increase the time required to optimize the sensor. This rapidly growing computational burden motivates the development of efficient surrogate modeling and intelligent optimization frameworks to accelerate both forward prediction and inverse design.

Machine learning (ML) has recently emerged as an effective approach to accelerating computational modeling and inverse design. Instead of repeatedly solving the governing electromagnetic equations, ML models learn the nonlinear relationship between structural parameters and sensor performance directly from simulation data. Once trained, these surrogate models can rapidly predict sensor characteristics, enabling efficient exploration of large design spaces while substantially reducing computational cost. A typical TMM simulation takes around 3-4 sec, while an ML model can predict the output parameters in milliseconds, providing an advantage of 3-4 orders of magnitude ($10^3$-$10^4$). Further details are in the Supplementary Information (Table S1). Consequently, ML has attracted considerable attention for forward prediction, optimization-assisted design, inverse design, and the development of intelligent photonic sensors [4][5].

Recent studies have demonstrated the successful application of Artificial Neural Networks (ANNs), ensemble learning methods, deep learning, and optimization algorithms for SPR and photonic sensor design. Graphene-assisted tunable SPR sensors optimized using ML and Particle Swarm Optimization (PSO) have achieved enhanced sensing performance with significantly reduced computational cost [6]. ANN-based photonic crystal fiber sensors have demonstrated accurate prediction of sensing characteristics [7], whereas boosting algorithms such as XGBoost and CatBoost, combined with SHAP analysis, have improved both prediction accuracy and model interpretability [8]. ML has also been successfully applied to fiber-optic-based dengue detection [9], terahertz biosensor optimization using Random Forest Regression and PSO [10][11], Bayesian Regularized Neural Networks for cancer-cell detection [12], multi-channel SPR sensing using CatBoost [13], and hemoglobin monitoring using locally weighted regression [14]. Furthermore, explainable artificial intelligence techniques, particularly SHAP and LIME, have enabled a better understanding of feature importance in SPR sensor optimization [15]. Hybrid optimization frameworks combining ANN, Genetic Algorithms, and Taguchi methods [16], sequence-based deep learning models for rapid sensor prediction [17], Random Forest classifiers for DNA detection [18], and machine-learning-assisted processing of Tilted Fiber Bragg Grating spectra [19] collectively demonstrate the growing impact of artificial intelligence in optical sensing.

Even with this progress, several significant challenges persist. Most reported studies address only a single aspect of the design process, such as forward prediction, optimization, or explainability. In contrast, comprehensive frameworks integrating accurate surrogate modeling, inverse design, explainable AI, robustness assessment, and independent validation remain limited. Moreover, systematic benchmarking of multiple state-of-the-art regression models, perturbation-based robustness analysis, and parameter recovery validation have received relatively little attention for multilayer prism-based SPR sensors operating in the near-infrared region.

Motivated by these limitations, this work presents a comprehensive ML-assisted framework for the analysis and inverse design of multilayer prism-based SPR sensors operating at 1550 nm. A large simulation dataset was generated using the TMM method by systematically varying material properties and structural parameters of multilayer SPR configurations. Using this dataset, multiple state-of-the-art regression models, including Multi-Layer Perceptron (MLP), XGBoost, LightGBM, and CatBoost, were trained and benchmarked to predict key sensing characteristics simultaneously. Model interpretability was investigated using SHAP analysis, while perturbation-based robustness analysis was conducted to assess prediction stability under input uncertainty. Furthermore, multiple optimization algorithms were integrated with the trained surrogate models to perform inverse design of multilayer SPR sensors. The optimized configurations were subsequently verified through parameter recovery analysis, thereby establishing the physical consistency and reliability of the proposed framework. Overall, the proposed framework significantly reduces the computational burden associated with conventional TMM-based design while providing a reliable and efficient approach for both forward prediction and inverse design of multilayer SPR sensors. In addition, the robustness of the optimization outcome was independently confirmed through a complementary forward design task, in which multiple optimizers converged on the same optimal configuration, which was directly validated against TMM simulation.

## II. Methodology & Work Flow

### A. *Performance parameters for SPR sensors:*

SPR sensors are usually studied and compared based on sensitivity, detection accuracy (DA), and figure of merit (FOM), which are the most insightful parameters for analyzing sensor performance [2].

Under an angle interrogation scheme, sensitivity (S) is defined as the ratio of change in resonance angle $(\Delta\theta_{res})$ with respect to change in the respective analyte's refractive index $(\Delta n_a)$

$$S = \frac{\Delta\theta_{res}}{\Delta n_a} (°/RIU) \tag{2}$$

The full width at half maximum (FWHM) is defined as the change or difference in resonance angles at 50% reflected intensity. FWHM intuitively defines how broad or narrow the SPR curves are. FWHM can also be used to calculate the detection accuracy, which is the reciprocal of FWHM.

$$DA = \frac{1}{FWHM} degree^{-1} \tag{3}$$

FOM is defined as the ratio of sensitivity to FWHM and provides a complete picture of sensor performance; it is also the most crucial parameter to seek.

$$FOM = \frac{S}{FWHM} (RIU^{-1}) \tag{4}$$

These parameters are calculated from the sensor configuration's reflection characteristics using the TMM method, as shown in Figure 2, which is a conventional approach for studying the optical characteristics of an n-layer stacked structure [3].

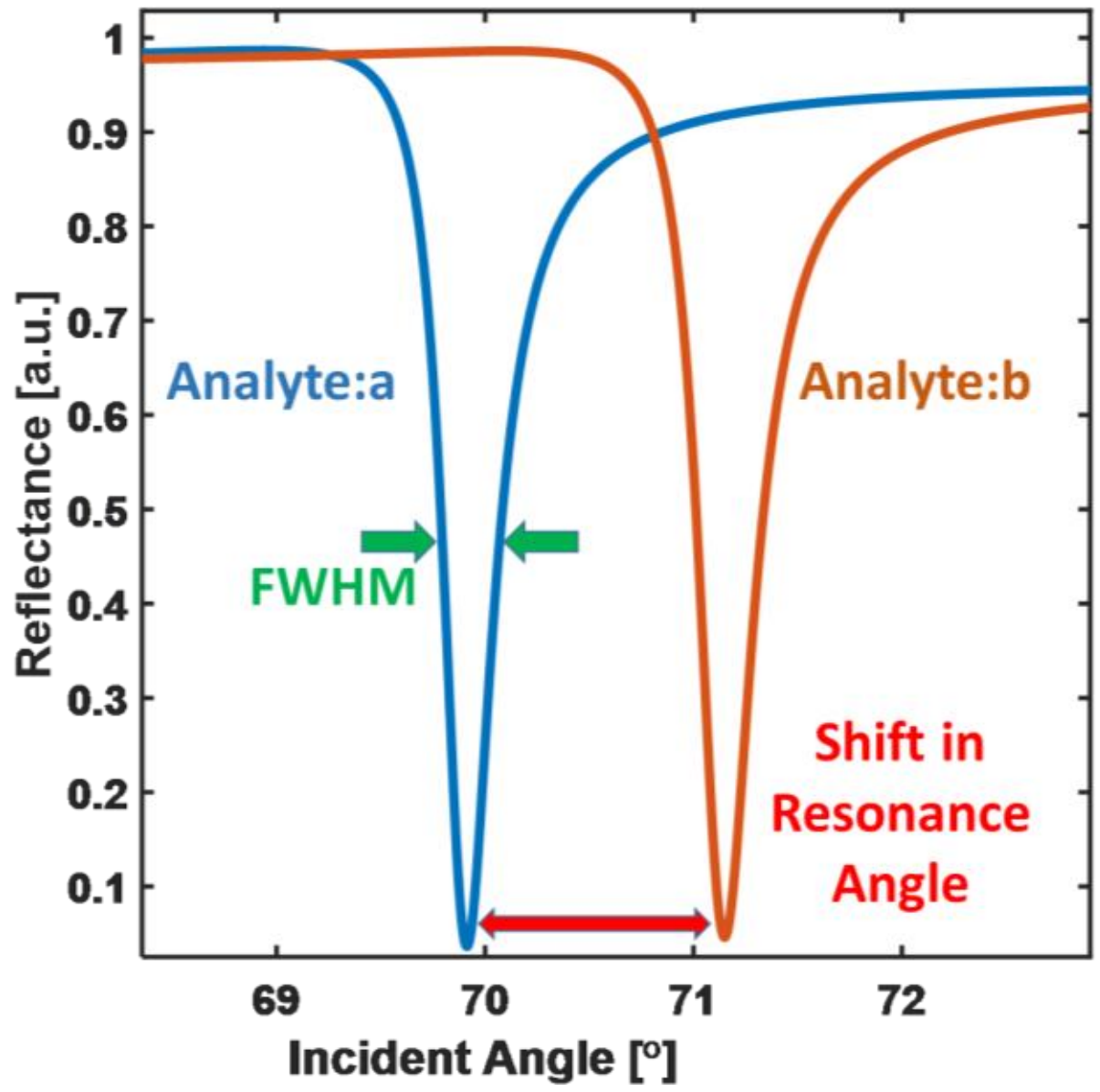

**Fig. 2**. Pictorial depiction for measurement of performance parameters using simulated reflectivity curves.

### B. *Machine Learning Performance Metrics:*

$R^2$ (Coefficient of Determination) measures how much of the variation in the true data is explained by the model, and values closer to 1 indicate a better fit. A higher $R^2$ means the model captures the underlying trend more effectively.

$$R^2 = 1 - \frac{\sum_{i=1}^{n} (y_i - \hat{y}_i)^2}{\sum_{i=1}^{n} (y_i - \bar{y})^2} \tag{5}$$

RMSE (Root Mean Squared Error) is the square root of the average squared prediction error, so it reflects the typical magnitude of the error in the same units as the output. Because larger errors are squared, RMSE penalizes big deviations more strongly than small ones, so lower values indicate better performance.

$$\text{RMSE} = \sqrt{\frac{1}{n}\sum_{i=1}^{n} (y_i - \hat{y}_i)^2} \tag{6}$$

MAE (Mean Absolute Error) is the average of the absolute differences between predicted and actual values, which makes it easy to interpret as the average prediction error.

$$\text{MAE} = \frac{1}{n}\sum_{i=1}^{n} |y_i - \hat{y}_i| \tag{7}$$

Here $y_i$ = actual (true) value for observation $i$, $\hat{y}_i$ = predicted value for observation $i$, $\bar{y}$ = mean of actual values $(\frac{1}{n}\sum_{i=1}^{n} y_i)$, and $n$ = number of observations. Unlike RMSE,

MAE assigns equal weight to all errors, making it less sensitive to outliers, and smaller MAE values indicate better accuracy.

Joint accuracy is the proportion of samples for which the predicted joint outcome (simultaneously high FOM and low $R_{min}$) matches the actual outcome. It summarizes overall correctness across all samples, so higher values indicate better agreement between predicted and observed results. Here, the positive class is defined as configurations falling in the top 30% of FOM values and the bottom 30% of $R_{min}$ values, i.e., physically desirable, high-performance sensor designs. Joint precision measures the fraction of samples predicted as positive that are true positives; a higher precision means fewer false alarms (configurations wrongly flagged as high-performance). Joint recall measures the fraction of actual positive samples that the model successfully identifies; a higher recall means fewer missed high-performance configurations. Both precision and recall are important in this context, since false positives could misdirect subsequent inverse design toward suboptimal regions, while false negatives could cause promising configurations to be overlooked.

The model's performance was evaluated using $R^2$, RMSE, MAE, joint accuracy, joint precision, joint recall, and correlation. Joint accuracy, precision, and recall assess the correctness of combined predictions, especially overall accuracy, positive predictive reliability, and detection sensitivity. Correlation indicates the strength of agreement between predicted and observed values; values closer to 1 indicate a stronger linear association.

The overall workflow of the proposed framework is illustrated in Figure 3. A simulation dataset was generated by systematically varying the multilayer SPR sensor parameters within predefined design ranges. The generated dataset was subsequently used to train, validate, and test the machine learning models, followed by inverse design using the selected surrogate model; detailed characteristics and statistical analysis are presented in Section III. The dataset was divided into training and testing subsets, and five-fold cross-validation was employed during model development to ensure a robust evaluation of performance. Multiple machine learning regression models, namely XGBoost, LightGBM, CatBoost, and MLP, were selected to represent complementary learning paradigms. XGBoost, LightGBM, and CatBoost are widely adopted gradient boosting algorithms well-suited for structured tabular datasets and capable of modeling complex nonlinear relationships. In contrast, MLP was included as a representative neural network-based approach for nonlinear function approximation. The trained models were evaluated comparatively using multiple statistical performance metrics, computational efficiency, and generalization capability to identify the most suitable surrogate model for predicting the performance of multilayer SPR sensors.

Their robustness was further evaluated using SHAP-guided perturbation analysis, in which controlled perturbations were applied to the input features to assess model stability under varying noise levels, and learning curve analysis was employed to examine convergence behavior. In addition, SHAP-based explainability analysis was performed to quantify the contribution of individual input parameters to the model predictions. To evaluate the robustness of the trained machine learning models against fabrication-induced uncertainties, a SHAP-weighted perturbation framework was developed. Eight SPR design parameters, including the prism refractive index ($n_c$), metal refractive index components ($M_{_re}$ and $M_{_im}$), metal thickness ($t_{Mt}$), dielectric refractive index ($n_{Di}$), dielectric thickness ($t_{Di}$), and the refractive index ($n_{2d}$) and thickness ($t_{2d}$) of the 2D material, were perturbed using zero-mean Gaussian noise. Additive perturbations were applied to absolute-valued parameters ($n_c$, $t_{Mt}$, $t_{Di}$, and $t_{2d}$), whereas multiplicative percentage-based perturbations were used for refractive-index parameters ($M_{_re}$, $M_{_im}$, $n_{Di}$, and $n_{2d}$). After perturbation, all features were clipped to their predefined physical bounds to ensure that sensor configurations were physically realizable. The perturbation magnitude was weighted according to the normalized SHAP feature importance, assigning larger perturbations to more influential parameters. The effective perturbation intensity was progressively increased, and the trained models were evaluated at each perturbation level using $R^2$, RMSE, and MAE. This framework enables systematic assessment of model robustness under realistic fabrication tolerances and input uncertainties.

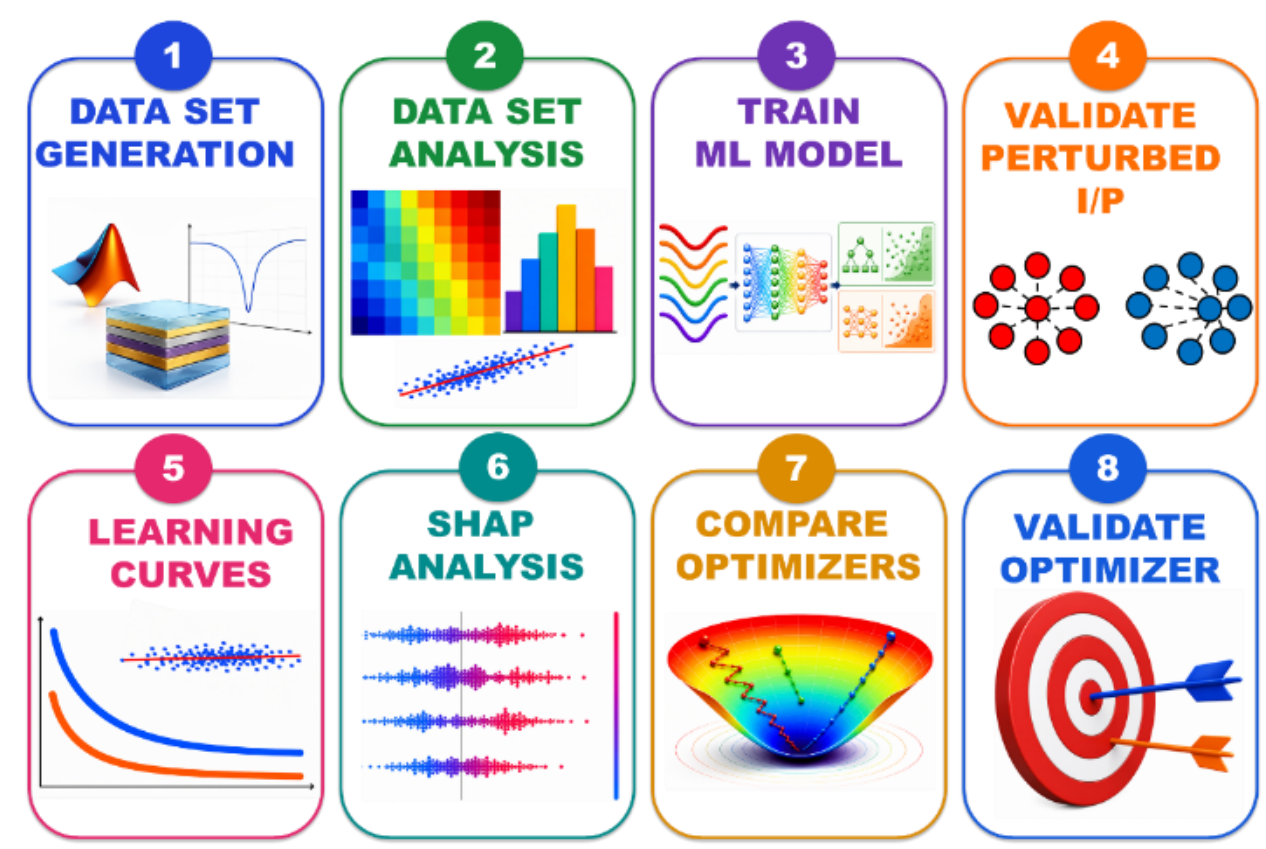

**Fig. 3.** A pictorial representation of the overall flow of work carried out.

The inverse-design optimization was formulated to identify sensor configurations that maximize the FOM while minimizing the $R_{min}$. Based on the comparative evaluation, the selected surrogate model was subsequently integrated with four complementary global optimization algorithms, namely differential evolution (DE), dual annealing (DA), DIRECT, and Bayesian Gaussian Processes (BGP), to perform inverse design. These optimizers were selected because they represent evolutionary, stochastic annealing, deterministic partitioning, and surrogate-assisted optimization strategies, enabling a comprehensive evaluation of optimization performance. Finally, the recovered sensor parameters and their corresponding performance metrics were validated against the target values to assess the reliability of the proposed inverse-design framework.

## III. Results and Analysis

### A. *Data set generation and Analysis:*

A large simulation-driven dataset was generated using the TMM by systematically varying the optical and structural parameters of multilayer SPR sensors operating in the Kretschmann configuration at 1550 nm. To ensure broad coverage of the design space while maintaining physical relevance, a generalized multilayer structure consisting of Prism/Metal/Dielectric/2D Material/Sensing Medium was considered. The materials were selected from commonly reported SPR sensor-based literature. BK7 and $CaF_2$ were used as prism materials; Au, Ag, and Al as plasmonic metals; $SiO_2$, $TiO_2$, and Si as dielectric layers; and Antimonene (Sb), fluorinated graphene (FG), and $MoS_2$ as representative 2D materials. These materials span a wide range of refractive indices and optical properties, enabling the dataset to capture diverse SPR responses. Combining the selected materials resulted in 54 unique multilayer architectures (2 × 3 × 3 × 3). For each architecture, metal and dielectric thicknesses were varied over physically realistic ranges in 1-nm increments, while the 2D material thickness was varied in monolayer increments. The analyte refractive index was varied between 1.33 and 1.34. For every configuration, the reflectance characteristics were computed using TMM, and the corresponding resonance angle, minimum reflectance ($R_{min}$), sensitivity, FWHM, and FOM were extracted. This process generated approximately $1.5 \times 10^5$ unique sensor configurations. The parameter ranges were selected based on commonly reported values in the SPR literature to ensure physically realizable sensor configurations and to provide sufficient diversity for machine learning training. The refractive indices and thicknesses of the constituent layers were used as input features, whereas the SPR performance parameters served as output variables. The generated dataset contains both high- and low-performing sensor configurations, providing sufficient diversity for supervised learning, explainable AI analysis, and optimization-assisted SPR sensor design.

TABLE II: List of materials, RI, and thickness ranges used to generate the data set

| Material/ Layer in stack | Name | RI | Thickness Range |
|---|---|---|---|
| Prism | BK7<br>$CaF_2$ | 1.5007<br>1.426 | -- |
| Metal | Gold (Au)<br>Silver (Ag)<br>Aluminum(Al) | 0.5747-i9.6643<br>0.3996-i10.173<br>0.4598-i14.4882 | 30nm-60nm<br>25nm-60nm<br>25nm-60nm |
| Dielectric | $SiO_2$<br>$TiO_2$<br>Si | 1.4596<br>2.4328<br>3.46971 | 1-10 nm |
| 2D Nanomaterial | Sb<br>FG<br>$MoS_2$ | 1.696<br>2.62563 -i1e$^{-6}$<br>3.647 | 1-10 layers |

To ensure reproducibility of the dataset generation process, Table II summarizes all materials considered in this study along with their refractive indices and the corresponding thickness ranges used in the TMM simulations. The dataset was generated through systematic parameter sweeps over all feasible combinations of materials and thicknesses within the predefined ranges. The generated dataset was randomly divided into training (80%) and testing (20%) subsets, while five-fold cross-validation (CV) was employed during model development to assess generalization performance.

The input variables were analyzed using a correlation matrix to assess multicollinearity among the structural and optical parameters. As shown in Figure 4, all pairwise correlations were low to moderate, with the highest correlations observed between the real and imaginary parts of the metal refractive index ($M__{re}$ and $M__{im}$, 0.32) and between the 2D material refractive index and thickness ($n_{2d}$ and $t_{2d}$, -0.29). Since no strong multicollinearity was observed, all input features were retained for subsequent model training. While interfeature correlations remained weak, several input parameters exhibited strong correlations with the output variables, indicating their significant influence on SPR performance. In particular, the imaginary part of the metal refractive index ($M__{im}$) showed a strong negative correlation with FOM (-0.90), highlighting the dominant role of metallic losses in determining resonance quality and sensing performance. Similarly, the metal thickness ($t_{Mt}$) exhibited a relatively strong positive correlation with the $R_{min}$ (0.59), indicating its influence on resonance depth. Furthermore, the positive correlation between FOM and $R_{min}$ (0.75) reflects the intrinsic coupling between resonance sharpness and reflectance characteristics in multilayer SPR structures. Overall, the observed correlation trends are consistent with established SPR physics and indicate that the selected structural and optical parameters provide meaningful predictive information for machine learning-based modeling and inverse design.

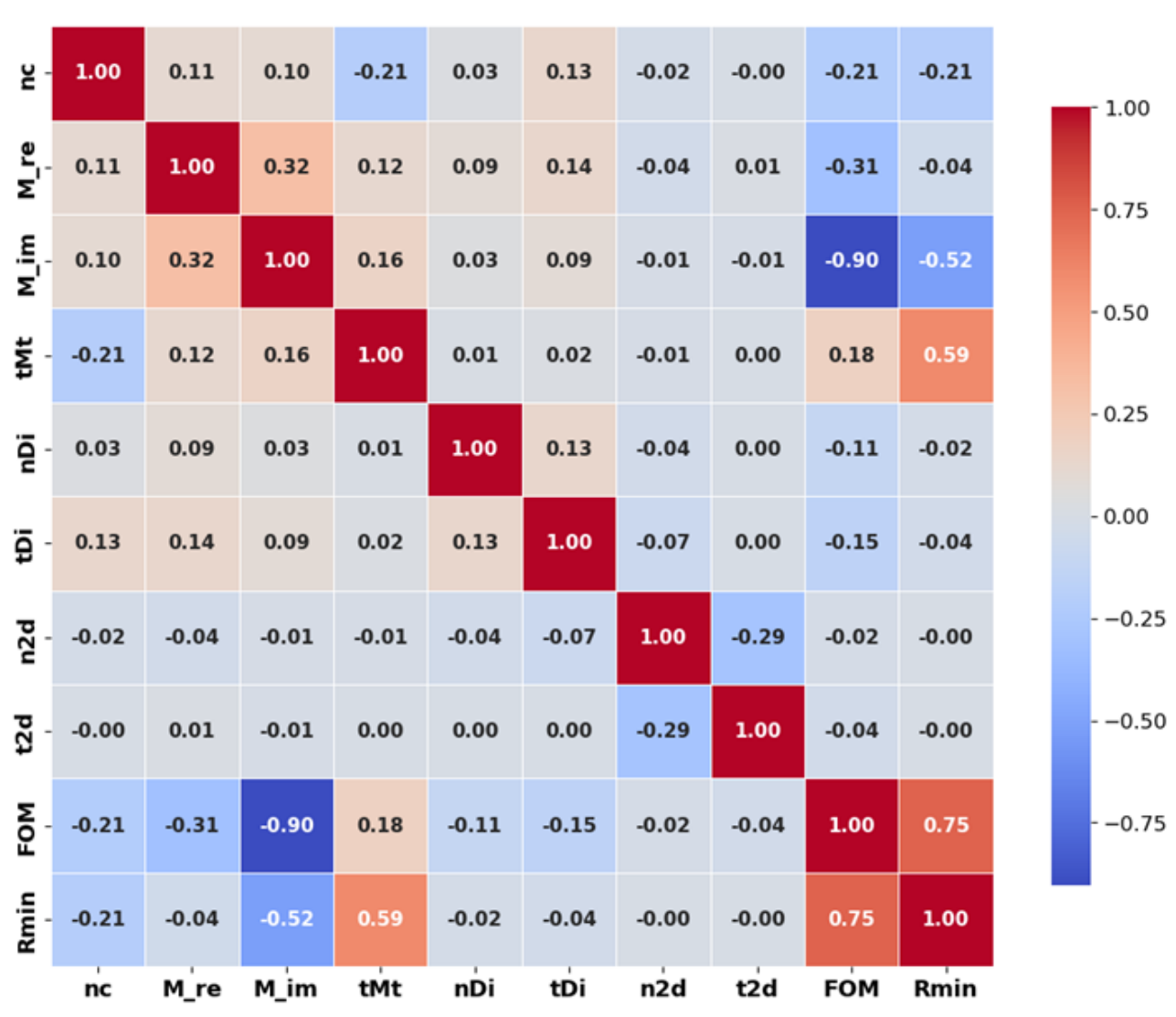

**Fig. 4.** Correlation matrix of input variables and output variable, used to assess relationships and multicollinearity among features.

Since the target variables, FOM, $R_{min}$, sensitivity, and resonance angle are all continuous numerical quantities, the problem is formulated as a supervised regression task. The generated dataset exhibits strong nonlinear coupling among material properties, layer thicknesses, refractive indices, and SPR responses, necessitating regression models capable of learning complex multidimensional relationships. Accordingly, four nonlinear regression models, namely MLP, XGBoost, LightGBM, and CatBoost, were selected for comparative evaluation. MLP represents a neural network-based approach capable of approximating complex nonlinear physical relationships. In contrast, XGBoost, LightGBM, and CatBoost are state-of-the-art gradient boosting algorithms that have demonstrated excellent performance on structured numerical datasets. These boosting methods efficiently capture nonlinear feature interactions, require minimal feature engineering, and offer high computational efficiency for large simulation-generated datasets. Together, the selected models provide complementary learning strategies to accurately predict SPR performance parameters and identify the most suitable surrogate model for subsequent inverse-design optimization.

### B. *Training and Comparative Analysis of Different ML models*

An optimal SPR sensor should simultaneously exhibit a high FOM and a low $R_{min}$. Accordingly, the predictive performance of the investigated machine learning models was first evaluated individually for each output parameter and subsequently through a joint analysis to assess their ability to identify high-performance sensor configurations.

Fig. 5 shows the True and predicted values for FOM and $R_{min}$ of the four models considered. The results indicate that all four machine learning models exhibit excellent agreement between the predicted and true FOM values, demonstrating high prediction accuracy. In contrast, the prediction of $R_{min}$ shows relatively greater variation than the FOM predictions. Among the models, LightGBM and CatBoost achieve the highest prediction accuracy, with minimal scatter around the ideal diagonal, while XGBoost exhibits noticeable deviations in the range of 0.2-0.5. The MLP model shows the largest dispersion and comparatively poorer prediction performance for $R_{min}$, indicating that ensemble-based methods provide more robust and reliable predictions for this parameter. Further hyperparameters are mentioned in the SI (Table S2).
Tables III and IV summarize the regression performance of the four investigated models for predicting FOM and $R_{min}$. Overall, all models achieved excellent predictive accuracy for FOM, with $R^2$ exceeding 0.999, demonstrating their ability to capture the nonlinear relationship between multilayer structural parameters and SPR performance. Among the models, CatBoost achieved the highest prediction accuracy for FOM, yielding the lowest RMSE (4.48) and MAE (2.31). For $R_{min}$ prediction, larger performance differences were observed. LightGBM achieved the best overall performance with an $R^2$ of 0.9957, RMSE of 0.0577, and MAE of 0.0121, whereas MLP exhibited comparatively lower accuracy ($R^2 \approx 0.9575$) and higher prediction error as shown in Table IV. These results indicate that the boosting-based models consistently outperform the neural network baseline in predicting multilayer SPR responses.

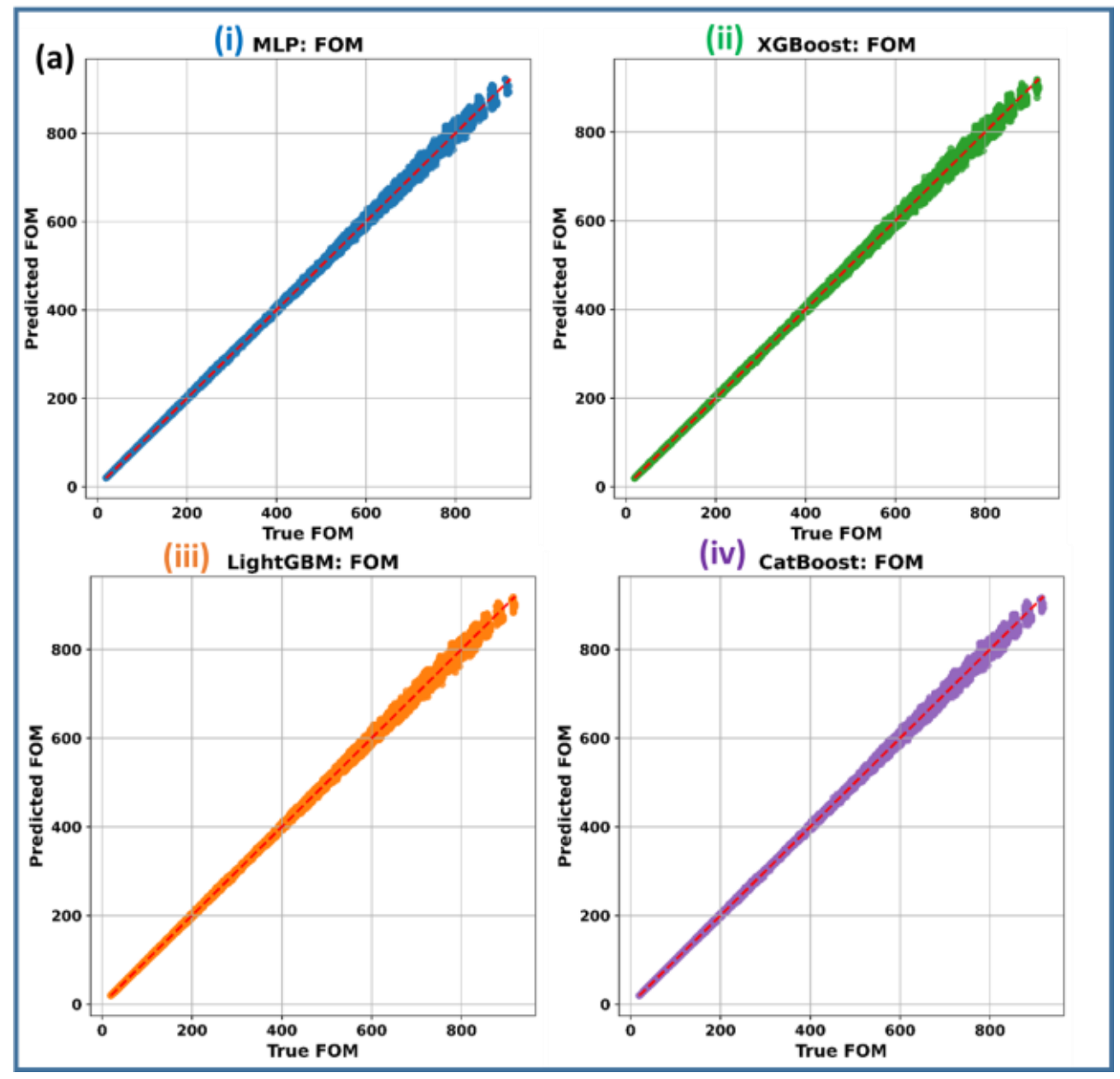

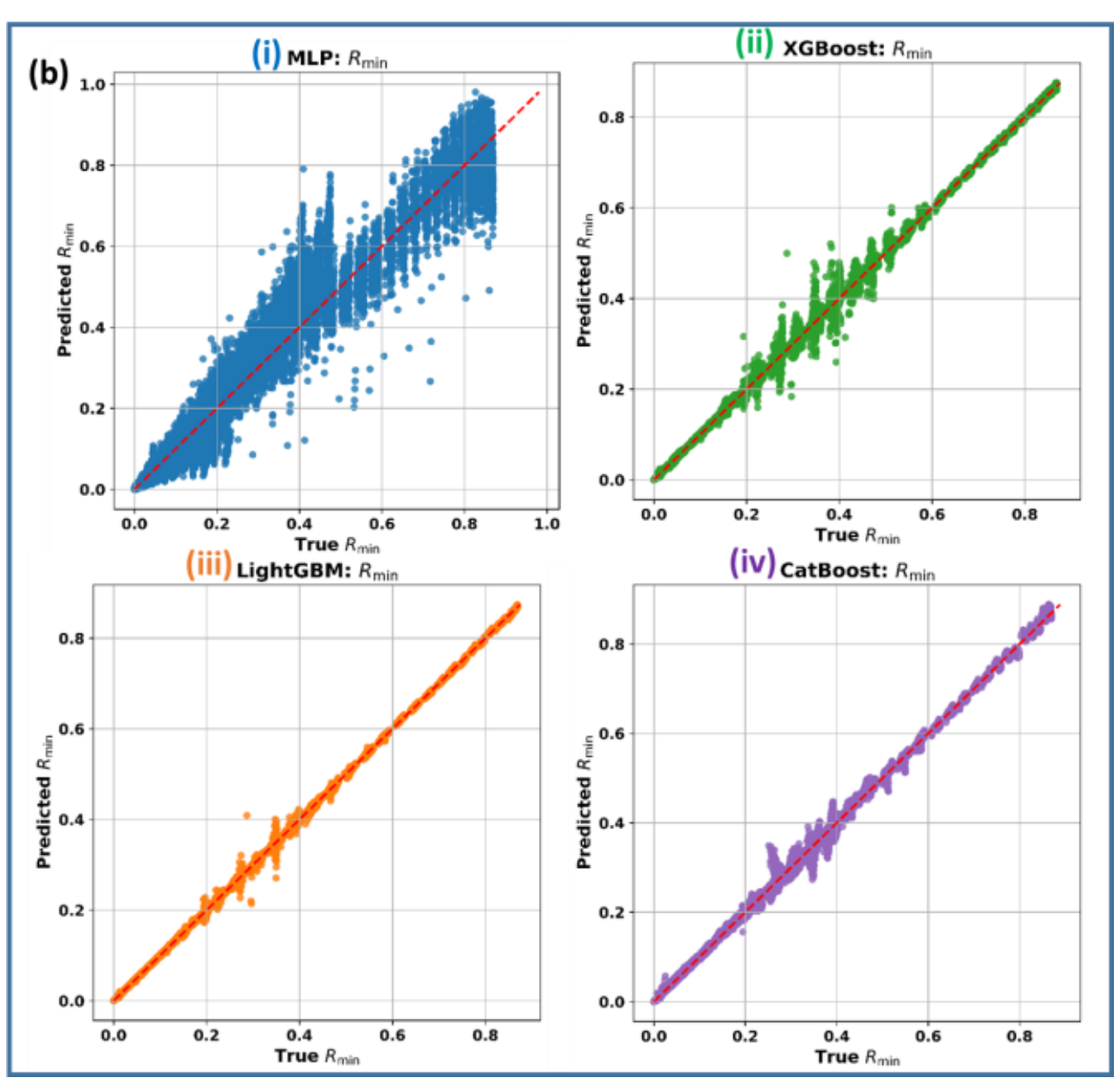

**Fig. 5**. True vs. Predicted Plots for (a) FOM and (b) $R_{min}$ of the four analyzed ML Models.

TABLE III: Comparative analysis of different ML models for FOM

| Model (FOM) | $R^2$ | RMSE | MAE |
|---|---|---|---|
| MLP | 0.999605 | 4.695693349 | 2.650010227 |
| XGBoost | 0.999605 | 4.698639259 | 2.499315697 |
| LightGBM | 0.999618 | 4.619579479 | 2.454431142 |
| CatBoost | 0.999641 | 4.481293408 | 2.312262786 |

TABLE IV: Comparative analysis of different ML models for $R_{min}$

| Model ($R_{min}$) | $R^2$ | RMSE | MAE |
|---|---|---|---|
| MLP | 0.957500052 | 0.181247227 | 0.109694126 |
| XGBoost | 0.994588023 | 0.064677788 | 0.017077025 |
| LightGBM | 0.995700224 | 0.057650101 | 0.012093506 |
| CatBoost | 0.990594328 | 0.085265252 | 0.02524995 |

Since practical SPR sensor design requires the simultaneous optimization of FOM and $R_{min}$, regression accuracy alone is insufficient to evaluate model performance. Therefore, an additional joint evaluation was performed by defining desirable sensor configurations as those that simultaneously fall in the top 30% of FOM values and the bottom 30% of $R_{min}$ values. Joint accuracy, Joint precision, and Joint recall were then computed by comparing the predicted and actual high-performance configurations, providing an integrated assessment of each model's ability to identify physically favorable sensor designs.

TABLE V: Joint parameters analysis of different ML models

| Model | Joint Accuracy | Joint Precision | Joint Recall |
|---|---|---|---|
| MLP | 0.992293233 | 0.878955696 | 0.922757475 |
| XGBoost | 0.998245614 | 0.972817133 | 0.98089701 |
| LightGBM | 0.998903509 | 0.980279376 | 0.990863787 |
| CatBoost | 0.998370927 | 0.975247525 | 0.981727575 |

The results presented in Table V show that most models perform well in terms of joint accuracy. At the same time, MLP lags in joint precision and recall, indicating reduced reliability in identifying optimal sensor configurations. This is consistent with the trend noted in Section III.B, reinforcing the reduced reliability of MLP for multi-objective sensor identification. Beyond prediction accuracy, model generalization and computational efficiency are essential for inverse-design applications. Five-fold cross-validation was therefore employed to evaluate the stability of each model on unseen data. LightGBM, CatBoost, and XGBoost achieved a cross-validation score of 0.99, whereas MLP produced a bit lower score of 0.97.

To further evaluate model generalization beyond conventional regression metrics, learning curves were analyzed for both FOM and $\log(R_{min})$, as shown in Figure 6. For FOM prediction, all models exhibited rapid convergence and achieved validation $R^2$ values approaching unity, indicating excellent learning of the nonlinear relationships governing SPR performance with negligible overfitting. In the case of $\log(R_{min})$, slightly larger but still limited train-validation gaps were observed, reflecting the greater sensitivity of resonance minima to multilayer optical interactions. Among the investigated models, CatBoost exhibited the smallest train-validation gap, while LightGBM also demonstrated consistently stable convergence throughout training. Overall, the learning curves confirm that the investigated models successfully learned the underlying nonlinear SPR relationships rather than memorizing the training data, demonstrating good generalization capability. Learning curves were generated in the $\log(R_{min})$ because the regression models were trained in the transformed target space. Since $R_{min}$ spans several orders of magnitude and exhibits a highly skewed distribution near zero, logarithmic transformation improves numerical stability, learning efficiency, and model convergence. For visualization and physical interpretation, the predicted values were transformed back to the original $R_{min}$ scale in the true-versus-predicted plots, allowing direct comparison with the actual SPR response.

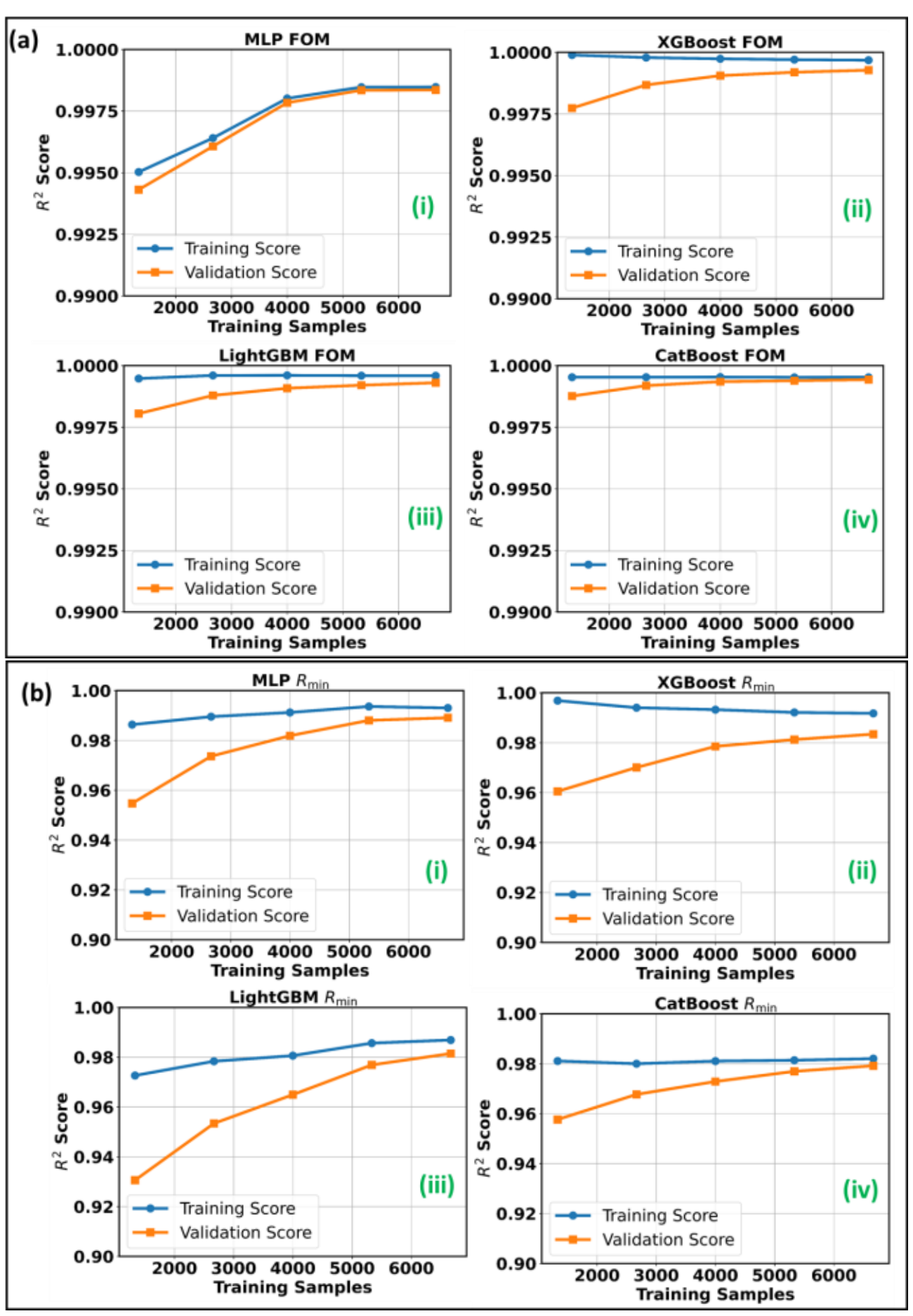

**Fig. 6**. Learning Curves for (a) FOM and (b) $R_{min}$ of the four analyzed models.

## C. *Feature Importance Analysis:*

SHAP (SHapley Additive exPlanations) is an explainable AI technique based on cooperative game theory that quantifies the contribution of each input feature to the model's prediction. To improve the interpretability of the trained machine learning models, SHAP analysis was employed to quantify the contribution of each input feature to the predicted FOM and $R_{min}$. The input features include $n_c$, M_re, M_im, $t_{Mt}$, $n_{Di}$, $t_{Di}$, $n_{2d}$, and $t_{2d}$. For a given prediction, the SHAP value represents the marginal contribution of each feature to the model output and is computed as [21].

$$\phi_i(f,x) = \sum_{S\subseteq F\setminus\{i\}} \frac{|S|!(|F|-|S|-1)!}{|F|!}\left[f_{S\cup\{i\}}(x_{S\cup\{i\}}) - f_S(x_S)\right] \quad (8)$$

where: $\phi_i$ is the SHAP value of feature $i$, $F$ represents the complete feature set, $S$ denotes a subset of features excluding feature $i$, and $f(x)$ corresponds to the trained ML prediction function.
To quantify the overall influence of each parameter, the global importance of each parameter is obtained by averaging the absolute SHAP values over all samples, using the global importance equation:

$$I_i = \frac{1}{N}\sum_{k=1}^{N} |\phi_i^{(k)}| \tag{9}$$

where: $I_i$ is the overall importance of feature $i$, $N$ is the total number of test samples, $\phi_i^{(k)}$ is the SHAP value of feature $i$ for the $k^{th}$ sample.

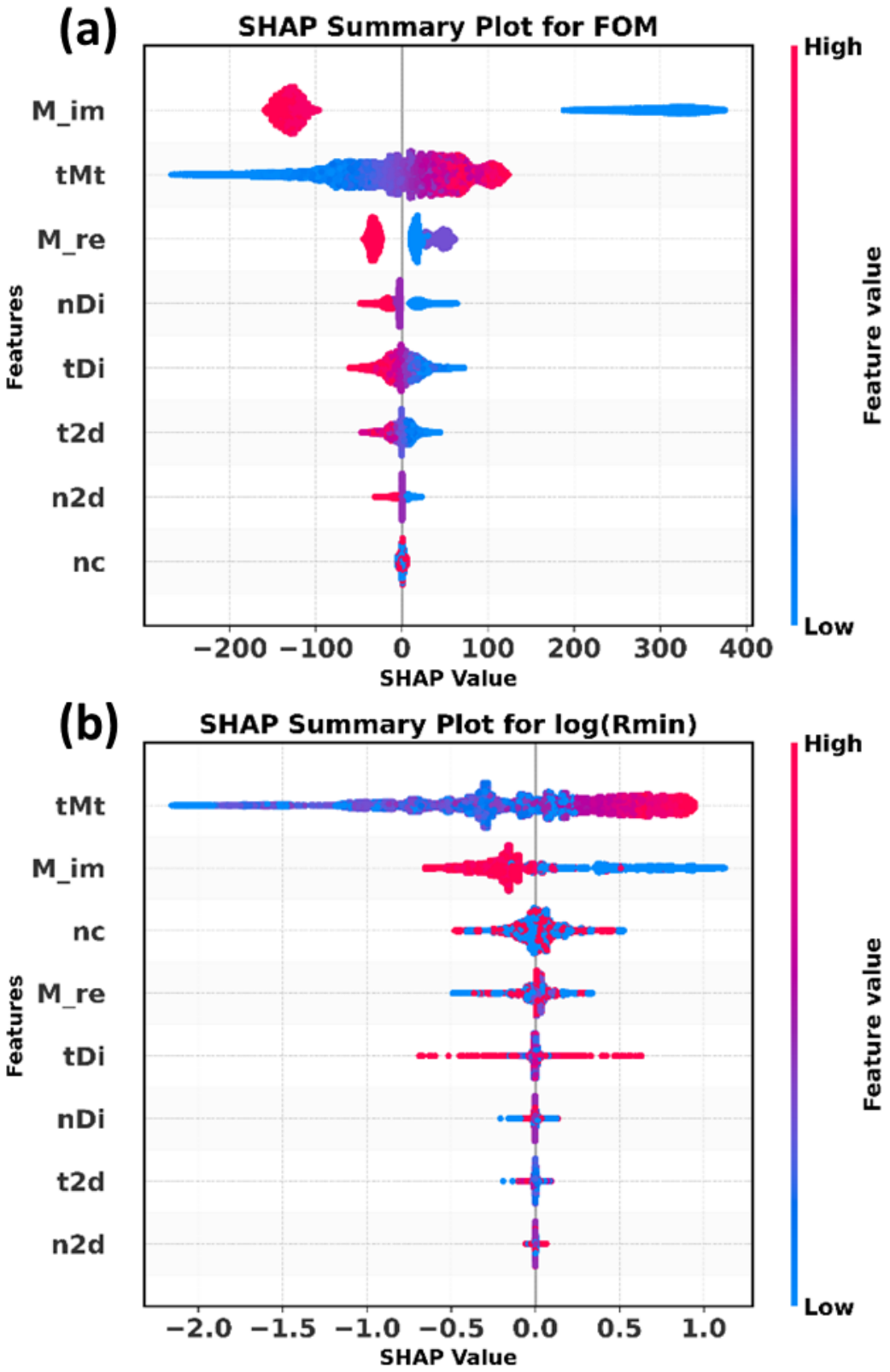

**Fig. 7.** SHAP feature analysis for the output parameters (a) FOM and (b) $R_{min}$, respectively.

The resulting SHAP summary plots rank the eight input features by their global importance in predicting FOM and log($R_{min}$), as shown in Figure 7. The SHAP analysis (Figure 7) identifies $M_{im}$ as the dominant feature governing both FOM and log($R_{min}$). It contributes approximately 56 % of the total importance for FOM prediction and 31.6% for log($R_{min}$). This indicates that absorption and plasmon damping majorly govern the quality and sensing performance. Next is the $t_{Mt}$, which contributes about 18% to FOM prediction and 51% to log($R_{min}$). This stronger influence on log($R_{min}$) is consistent with the well-established dependence of resonance depth and plasmon excitation efficiency on the thickness of the metallic film. $M_{re}$ exhibits a moderate influence. In contrast, $n_{Di}$ and $t_{Di}$ primarily contribute by fine-tuning the resonant condition and optical coupling. The contributions of $n_{2d}$ and $t_{2d}$ remain comparatively small, suggesting that they mainly modulate, rather than dominate, the SPR response. In contrast, $n_c$ has the smallest contribution to FOM but a comparatively larger influence on log($R_{min}$), indicating that changes in the optical coupling conditions affect the resonance minimum more strongly than the overall sensing performance.

Overall, the SHAP analysis demonstrates that loss-related optical parameters, particularly the imaginary part of the metal refractive index and the metal thickness, dominate the ML predictions. In contrast, the dielectric and structural parameters provide secondary modulation of the SPR response. These feature-importance trends are consistent with established SPR physics, indicating that the trained ML models have successfully captured physically meaningful relationships between multilayer structural parameters and sensor performance.

*D. Perturbation & Robustness Analysis:*

Figure 8 presents the robustness of the trained ML models under normalized SHAP-weighted input perturbations for predicting FOM and $R_{min}$. The perturbation framework simulates realistic fabrication-induced variations in the multilayer SPR design parameters while preserving physically valid sensor configurations. As the perturbation increases, the prediction accuracy ($R^2$) gradually decreases for all four models.

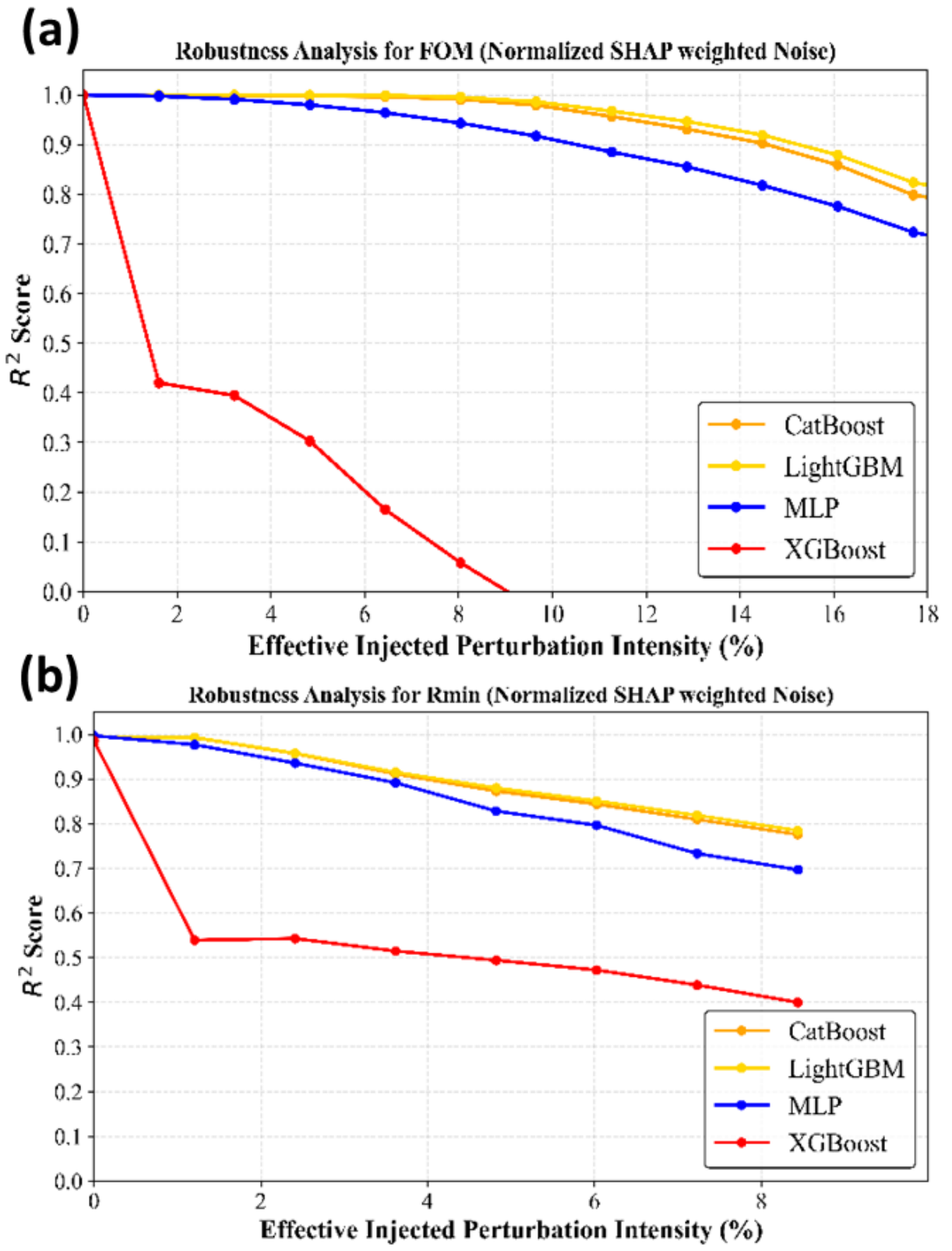

**Fig. 8.** Perturbation-based generalization analysis for (a) FOM and (b) $R_{min}$ prediction.

For FOM prediction (Figure 8(a)), CatBoost and LightGBM achieve high $R^2$ values within the defined perturbation range, demonstrating strong tolerance to variations in input parameters. In contrast, XGBoost exhibits a comparatively faster decline in prediction accuracy as the perturbation intensity increases, indicating lower robustness to input uncertainty, whereas MLP shows an intermediate level of performance. A similar trend is observed for $R_{min}$ prediction (Figure 8(b)), where the boosting-based models continue to outperform the neural network under perturbed conditions. The superior robustness of CatBoost and LightGBM can be attributed to their ability to capture complex nonlinear interactions among features while remaining less sensitive to moderate variations in structural and optical parameters. Since the applied perturbations emulate fabrication tolerances and measurement uncertainties commonly encountered in practical SPR sensor development, the results indicate that these models retain reliable predictive performance even under realistic operating conditions.

Overall, the perturbation analysis complements the regression, cross-validation, and learning-curve evaluations by demonstrating that the investigated models remain reliable beyond the nominal training distribution. This robustness ranking mirrors the regression and generalization trends reported above, further supporting CatBoost and LightGBM as the more reliable surrogates.

These results indicate that both boosting-based models are highly suitable surrogates for ML-assisted SPR inverse design, CatBoost showing an advantage in FOM accuracy (Table III), generalization stability (smallest train-validation gap), and robustness to input perturbation (Figure 8(a)), with LightGBM showing a marginal advantage in $R_{min}$ accuracy and joint-metric performance (Tables IV–V). As FOM is the dominant and controlling parameter in SPR design, the same will hold for inverse design tasks. CatBoost's combination of high FOM accuracy and strong generalization characteristics made it the preferred surrogate model for the inverse-design task.

## IV. OPTIMIZATION DRIVEN FORWARD AND INVERSE DESIGN ANALYSIS

The proposed inverse design framework uses the trained ML model for predicting the performance parameters of the SPR sensor. Instead of running many iterative TMM simulations directly during optimization, the trained model can estimate the FOM and $R_{min}$ much more quickly. This reduces the computational cost related to inverse design. For each layer, optical properties and bounds are similar to those used for data set generation. Four optimization algorithms, namely DE, DA, DIRECT, and BGP optimization, were employed to explore the design space and identify sensor configurations with improved optical performance. To ensure a fair comparison, all optimizers were evaluated using the same surrogate model and identical search space. The effectiveness of the optimization framework was assessed from three complementary perspectives. First, the convergence behavior of the optimization algorithms was analyzed to compare their search efficiency and stability. Subsequently, the inverse design capability was evaluated through parameter recovery and performance recovery analyses, in which the optimized sensor configurations were compared against target designs selected from the hidden test dataset. Finally, the consistency and reliability of the optimization outcome were examined through a complementary forward-design task, in which each optimizer's best-performing configuration was validated directly against a full TMM simulation.

### *A. Convergence Analysis of Optimization Algorithms:*

To compare the search efficiency of the optimization algorithms, the convergence behavior was analyzed by monitoring the evolution of the best objective value as function evaluations increased, shown in Figure 9. The objective value represents the optimization criterion used by the surrogate-assisted search framework, where lower values indicate improved optimization performance.

As illustrated in Figure 9, all four optimization algorithms exhibit a rapid decrease in the objective value during the initial stage of the optimization, indicating that the search is quickly directed toward promising regions of the design space. The inset for the first 100 function evaluations highlights early convergence, showing that DE achieves the fastest reduction in the objective value, followed by DA. BGP optimization also demonstrates rapid initial improvement, whereas DIRECT exhibits comparatively slower but more stable convergence. The global convergence history further reveals that the objective values gradually stabilize as the number of function evaluations increases, indicating successful convergence toward near-optimal solutions. Among the investigated methods, DE reaches the lowest objective value with the fewest evaluations, demonstrating an effective balance between global exploration and local exploitation. DA achieves a comparable final objective value with a slightly slower refinement phase, while DIRECT continues to improve steadily over a larger number of evaluations. BGP converges rapidly during the early iterations but reaches a stable solution at a comparatively higher objective value.

Overall, the convergence analysis demonstrates that all four optimization algorithms successfully identify feasible sensor configurations within the proposed machine learning-assisted optimization framework. However, DE exhibits the most efficient convergence behavior and achieves the best optimization performance, followed closely by DA. These observations are consistent with the analyses of parameter and performance recovery presented in the subsequent sections.

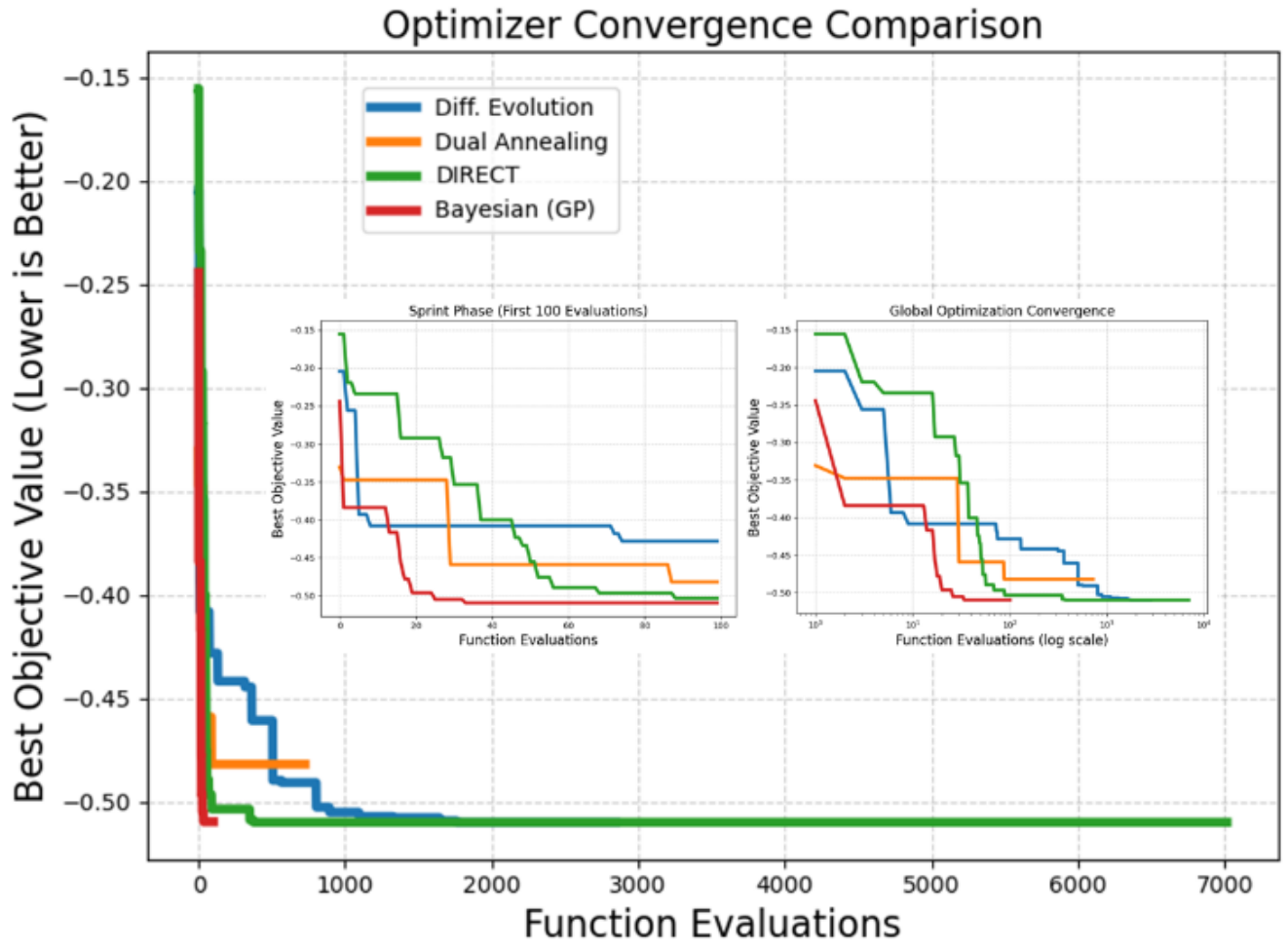

**Fig. 9.** Plots for convergence analysis of different optimization algorithms.

### B. *Inverse Design Performance Analysis:*

Following the convergence analysis, the inverse design capability of the proposed framework was evaluated by comparing the recovered sensor configurations with the corresponding target designs selected from the hidden test dataset. The assessment was performed from two complementary perspectives, namely the recovery of key design parameters and the reproduction of the desired optical performance. Since the SHAP analysis identified the metal optical constants as among the most influential parameters governing the SPR performance metrics, their recovery is presented as a representative assessment of the inverse-design capability. The inverse design performance was evaluated using 20 target sensor designs randomly selected from the test dataset. For each target, the optimization algorithms generated an optimal sensor configuration, and the recovery errors were averaged across all target designs. Figure 10 compares the average recovery error of the metal optical constants for the four optimization algorithms. The results indicate that all optimizers recovered the metal refractive index with relatively low error, although the accuracy of recovery varied between the real and imaginary components. BGP achieved the lowest error for the imaginary component, while exhibiting the highest error for the real component. DIRECT showed the largest error for the imaginary component (15.5%). DA and DE demonstrated a somewhat balanced performance, with moderate errors in both $M_{re}$ and $M_{im}$. Overall, the relatively low recovery errors indicate that the metal optical constants are reliably identified by the optimization algorithms, reflecting their dominant influence on the SPR resonance characteristics and confirming that these parameters are well constrained during the inverse design process. Other recovered parameters ($n_c$ and $t_{Mt}$) also showed a similar trend, as shown in the Supplementary Information (S3: Figure S1-S2). These findings highlight the inherent non-uniqueness of the inverse design problem, in which different combinations of material properties and layer thicknesses can yield nearly identical optical responses. Consequently, exact parameter recovery is not a prerequisite for achieving the desired sensor performance. Therefore, the optimized sensor configurations were further evaluated through a performance recovery analysis based on the target FOM and $R_{min}$.

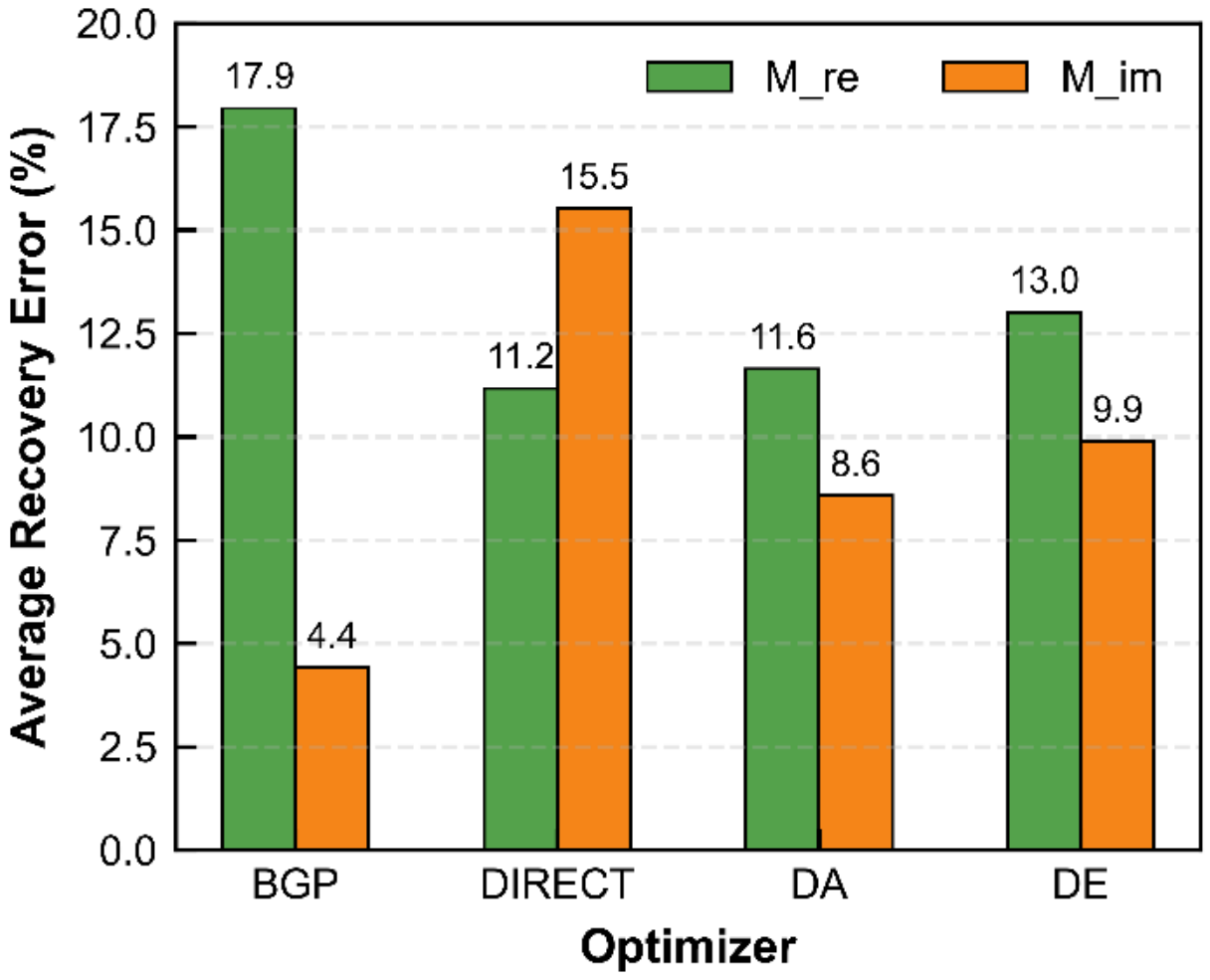

**Fig. 10.** Average recovery error of the metal RI for the four optimization algorithms.

Figure 11 compares the average recovery error of the target performance metrics, FOM and $R_{min}$, obtained using the four optimization algorithms. DE and DA achieved excellent recovery accuracy, with average errors of around 1% for both FOM and $R_{min}$, indicating an excellent ability to reproduce the desired sensor performance. In contrast, BGP and DIRECT exhibited comparatively larger deviations, with recovery errors ranging from approximately 7% to 9% for both performance metrics. Nevertheless, even these higher errors remain within an acceptable range for inverse design applications. The low recovery errors, particularly for DE and DA, demonstrate that the proposed machine learning-assisted optimization framework effectively identifies sensor configurations that reproduce the target optical characteristics.

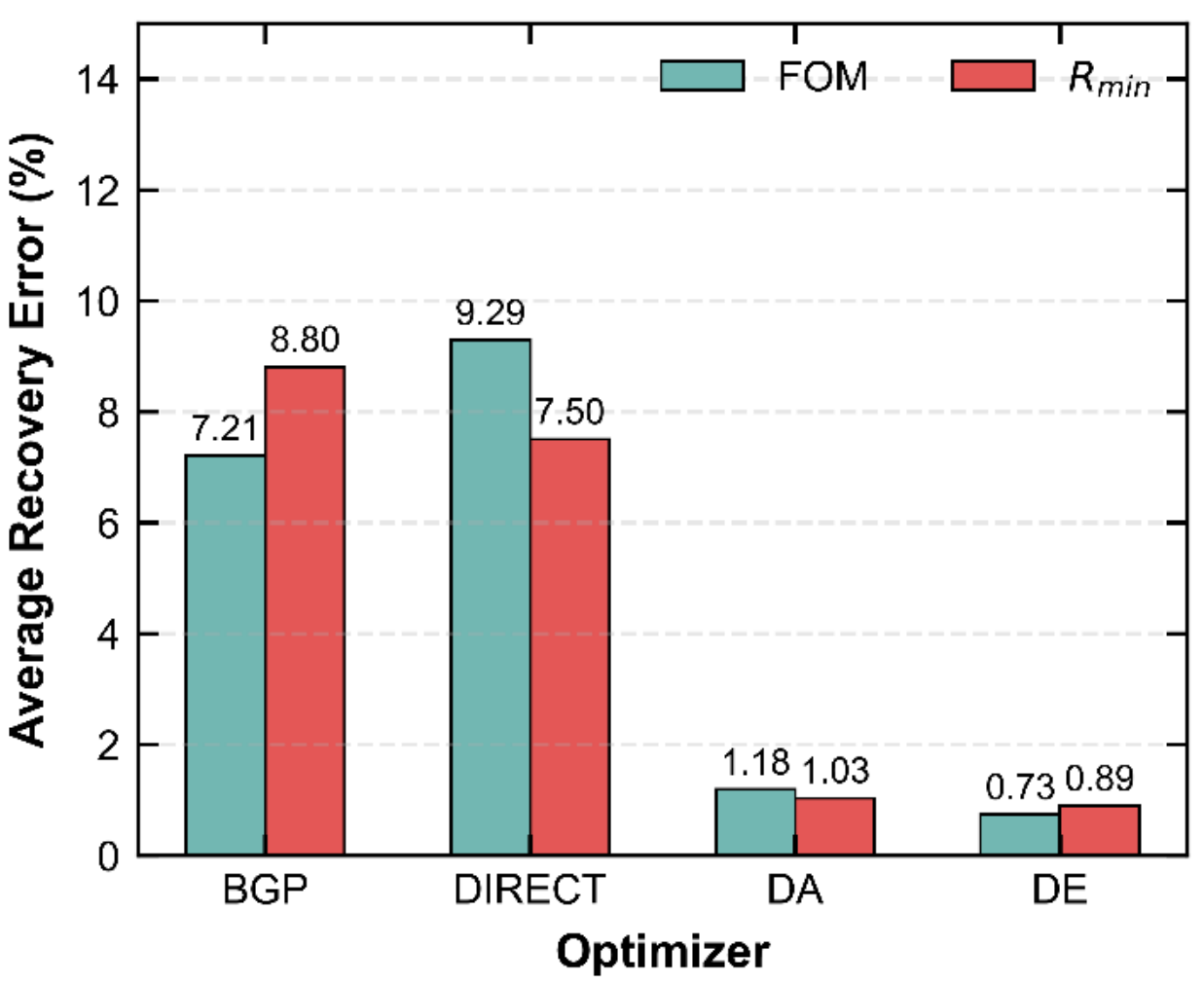

**Fig. 11.** Average recovery error of the performance parameters for the four optimization algorithms.

Overall, the proposed ML-assisted inverse design framework successfully identified SPR sensor configurations that closely reproduced the desired optical performance. Among the investigated optimization algorithms, DE demonstrated the best overall performance, followed closely by DA.

### *C. Forward Design Optimization- Cross-Optimizer Design Consensus and Direct TMM Validation:*

Sections A and B evaluated the optimizers on an *inverse design* task in which specific target sensor responses, selected from the hidden test dataset, must be recovered. This section examines the complementary *forward design* problem, in which each optimizer independently searches, without reference to any predefined target, for the single-sensor configuration that maximizes the predicted FOM-$R_{min}$ trade-off. To assess the robustness of this forward optimization outcome, each of the four optimization algorithms was executed 30 independent times, using different random seeds, over the discrete, simulation-validated design space. This discrete formulation restricts each optimizer to the set of layer thickness and refractive index combinations for which ground-truth TMM data exist, avoiding extrapolation of the CatBoost surrogate beyond its training distribution. Figure 12 shows the spread of each of the eight design parameters across all 30 runs for the four optimizers. The prism refractive index and M_re and M_im components of the metal refractive index remained essentially invariant across all runs and all optimizers (coefficient of variation < 0.01%), and the metal thickness varied only marginally (CV = 2.25%), indicating that the optimization tightly and consistently constrains these parameters. The dielectric and 2D-material layer parameters exhibited comparatively broader, but still bounded, variation, consistent with the inherent non-uniqueness of the inverse design problem discussed in Section B.

From these 30 runs, the single best configuration identified by each optimizer (lowest objective score, combining FOM and $R_{min}$) was extracted for direct comparison as shown in Table VI. Three of the four optimizers, DE, DA, and BGP, converged to the same optimal configuration across all eight design parameters, despite differing substantially in their search strategies and stochastic initializations. DIRECT converged to a configuration differing only in the 2D-material refractive index ($n_{2d}$ = 1.696 versus 2.626). Comparison of the achievable FOM at each of these two values (543.2 versus 553.5, a difference of 1.9%) indicates a shallow, broad optimum with respect to this parameter rather than a sharp single peak. However, the coarse discretization of $n_{2d}$ in the simulated database (three sampled values) does not entirely rule out a finer-resolution contribution in this region. This level of agreement across independent, algorithmically unrelated optimizers indicates that the identified optimum is a genuine, robust feature of the forward design landscape rather than an artifact of any single search algorithm.

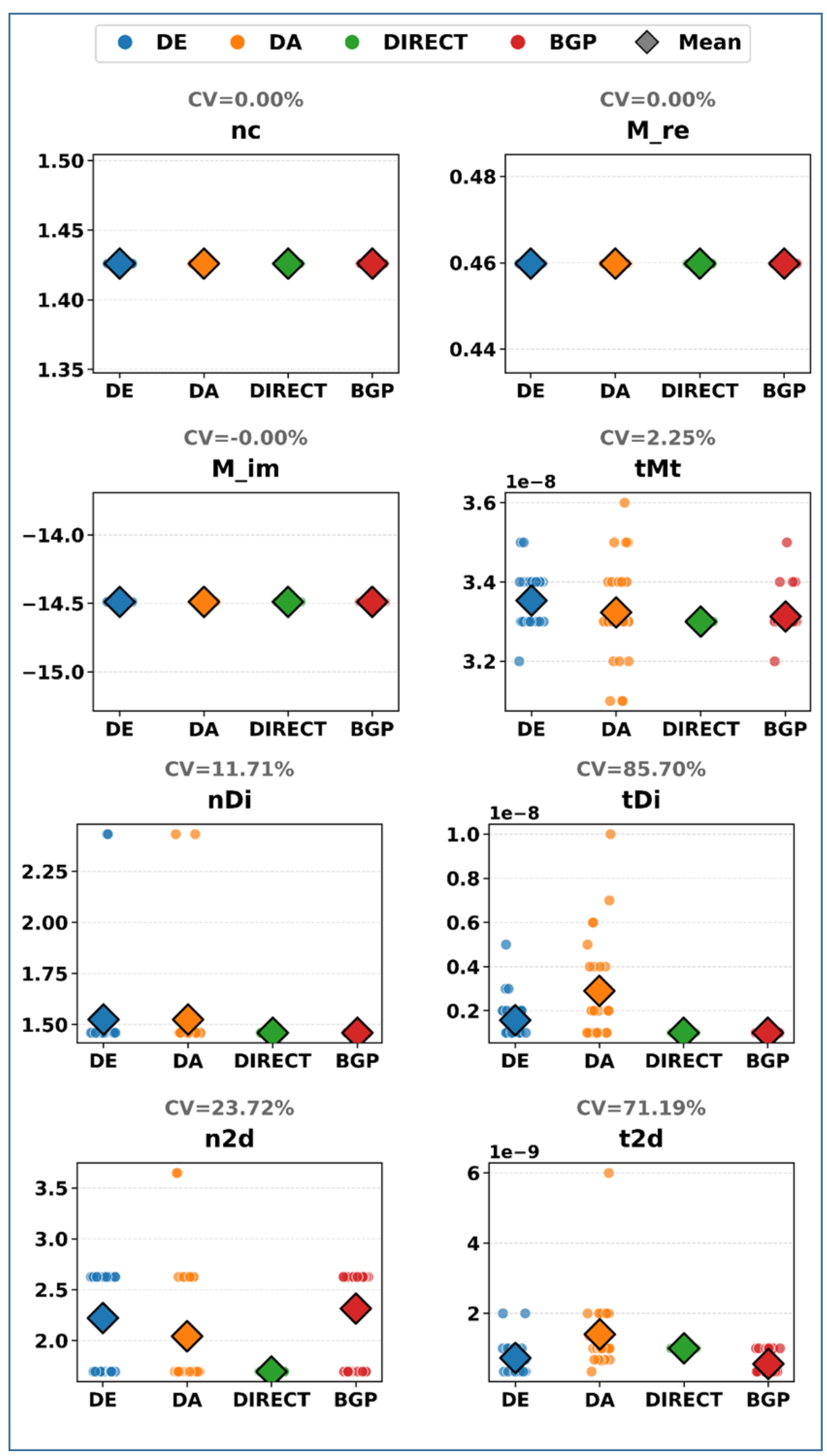

**Fig. 12.** Spread of each design parameter across 30 independent runs for the four optimization algorithms (forward design task).

The predicted performance of each optimizer's best configuration was subsequently validated through direct TMM simulation, rather than relying solely on the CatBoost surrogate. As summarized in Table VII, the ML-predicted FOM and $R_{min}$ matched the TMM-simulated ground truth to within 0.89% and 0.62% error, respectively, across all four optimizers, confirming that the surrogate-guided optimization framework identifies configurations whose predicted performance is physically accurate.

TABLE VI. Best design parameters identified by each optimizer (forward design task)

| Optimizer | DE | DA | DIRECT | BGP |
|---|---|---|---|---|
| nc | 1.426 | 1.426 | 1.426 | 1.426 |
| M_re | 0.4598 | 0.4598 | 0.4598 | 0.4598 |
| M_im | -14.49 | -14.49 | -14.49 | -14.49 |
| $t_{Mt}$ (nm) | 33 | 33 | 33 | 33 |
| $n_{Di}$ | 1.46 | 1.46 | 1.46 | 1.46 |

| $t_{Di}$ (nm) | 1 | 1 | 1 | 1 |
|---|---|---|---|---|
| $n_{2d}$ | 2.626 | 2.626 | 1.696 | 2.626 |
| $t_{2d}$ (nm) | 0.34 | 0.34 | 0.34 | 0.34 |

It should be noted that the low values reported in Table VII reflect the fidelity of the CatBoost surrogate in predicting the performance of a single, optimizer-selected configuration relative to direct TMM simulation, for the forward design task. These values are therefore not directly comparable to the target-recovery errors reported in Section B (7-9% for BGP and DIRECT), which instead quantify each optimizer's ability to reproduce a range of externally specified target responses in the inverse design task. The two metrics address distinct aspects of the optimization framework, predictive fidelity versus target-recovery accuracy, and do not represent conflicting results for the same quantity.

TABLE VII. TMM validation of the best configuration identified by each optimizer (forward design task)

| Optimizer | DE | DA | DIRECT | BGP |
|---|---|---|---|---|
| ML-FOM | 548.57 | 548.57 | 546.28 | 548.57 |
| TMM-FOM | 553.49 | 553.49 | 543.18 | 553.49 |
| FOM error(%) | 0.889 | 0.889 | 0.570 | 0.889 |
| ML- $R_{min}$ | 0.06617 | 0.06617 | 0.06597 | 0.06617 |
| TMM-$R_{min}$ | 0.06613 | 0.06613 | 0.06638 | 0.06613 |
| $R_{min}$ error (%) | 0.058 | 0.058 | 0.611 | 0.058 |

Taken together, the design consensus and TMM validation results demonstrate that, for the forward design task, the ML-assisted optimization framework not only converges to a consistent optimal sensor configuration across independent optimization algorithms, but that this configuration's predicted performance is confirmed by full-wave TMM simulation, reinforcing the reliability of the proposed surrogate-assisted optimization pipeline for practical SPR sensor design.

Further, a comprehensive comparison of the proposed framework with recent machine learning-based SPR sensor studies is presented in Table VIII. Although ML has been increasingly employed for the forward prediction and optimization of SPR sensor performance, studies addressing the inverse design of prism-based SPR sensors remain limited. Moreover, most existing works primarily focus on performance prediction or optimization, with comparatively less emphasis on model interpretability and inverse design assessment. In contrast, the present work integrates SHAP-based feature interpretation, perturbation-based robustness analysis, and a systematic inverse-design framework supported by parameter and performance recovery analyses. These complementary evaluations provide a more comprehensive assessment of the proposed ML-assisted SPR design framework than has typically been reported in the existing literature.

TABLE VIII: Comparative analysis of current work with recent works

| SHAP Analysis | Inverse Design | Learning Curves | Perturbation Analysis | Year |
|---|---|---|---|---|
| × | × | ✓ | × | 2025 [22] |
| × | × | × | × | 2026 [23] |
| × | × | × | ✓ | 2026 [24] |
| × | × | × | ✓ | 2026 [25] |
| × | × | × | × | 2026 [26] |
| × | ✓ | × | × | 2025 [27] |
| × | ✓ | × | × | 2024 [28] |
| ✓ | ✓ | ✓ | ✓ | Present Work |

## V. CONCLUSIONS

In this study, a machine learning-assisted framework was developed to analyze and design multilayer SPR sensor structures using a TMM-generated simulation dataset. Different regression models, including MLP, XGBoost, LightGBM, and CatBoost, were trained and evaluated to predict key sensing characteristics, such as FOM and $R_{min}$. Boosting-based models performed better than the other models analyzed at predicting performance parameters, thereby outperforming slower TMM simulations. The results show that the proposed ML framework effectively captures the complex nonlinear relationships among material properties, structural parameters, and SPR sensing performance. In addition, feature importance and SHAP-based interpretability analysis provided valuable insight into the parameters that most strongly influence sensor behavior and performance trends. The close agreement between ML-predicted and TMM-simulated outputs, together with cross-validation and perturbation-based generalization studies, confirms the reliability and robustness of the developed models. To further enhance sensor design efficiency, four optimization algorithms were integrated into the ML-assisted inverse-design pipeline to optimize the parameters of multilayer SPR structures. The comparative study demonstrated the optimizer's effectiveness in efficiently exploring the design space and identifying high-performance sensor configurations, with DA and DE showing similar performance. Notably, independent runs of three of the four optimization algorithms converged to an almost identical optimal sensor configuration in the forward design task. This configuration's predicted performance was confirmed by direct TMM simulation to within approximately 0.6-0.9% FOM error and under 0.7% $R_{min}$ error, demonstrating that the identified optimum is a robust feature of the design landscape rather than an artifact of any single search algorithm. However, rather than identifying a single universally best ML model or optimization algorithm, the present work primarily demonstrates the feasibility and effectiveness of ML-assisted SPR sensor design. This study shows that when ML models and optimization algorithms are properly trained, validated, and tuned on physically meaningful, diverse datasets, they can significantly accelerate exploration of complex SPR design spaces.

Future work will focus on expanding the physics-based dataset to improve model generalization across a wider range of SPR configurations. A limitation of the present framework is the coarse discretization of the $n_{2d}$ parameter (three candidate 2D materials), which does not entirely rule out a finer-resolution or continuous-material contribution to sensor performance. This is a natural direction for future refinement.